\documentclass[conference]{IEEEtran}
\usepackage{cite}
\usepackage{amsmath,amssymb,amsfonts}
\usepackage{algorithmic}
\usepackage{graphicx}
\usepackage{textcomp}
\usepackage{xcolor}
\usepackage{comment}
\usepackage{listings}
\usepackage{fancyvrb}

\usepackage{blkarray}
\usepackage{amsmath}
\usepackage{kbordermatrix}
\usepackage{dsfont}
\usepackage{amsthm}
\usepackage{lipsum}
\usepackage{graphicx}

\usepackage[utf8]{inputenc}
\usepackage[english]{babel}
\def\BibTeX{{\rm B\kern-.05em{\sc i\kern-.025em b}\kern-.08em
    T\kern-.1667em\lower.7ex\hbox{E}\kern-.125emX}}
\begin{document}

\title{\begin{small}\textcolor{blue}{\emph{© 2021 IEEE.  Personal use of this material is permitted.  Permission from IEEE must be obtained for all other uses, in any current or future media, including reprinting/republishing this material for advertising or promotional purposes, creating new collective works, for resale or redistribution to servers or lists, or reuse of any copyrighted component of this work in other works.}} \end{small}\\
A Non-intrusive Failure Prediction Mechanism for Deployed Optical Networks
\thanks{The research is funded by \emph{Tejas Networks}, Bangalore, India}
}

\author{\IEEEauthorblockN{Dibakar Das}
\IEEEauthorblockA{
\textit{IIIT Bangalore}\\
Bangalore, India \\
dibakard@ieee.org}
\and
\IEEEauthorblockN{Mohammad Fahad Imteyaz}
\IEEEauthorblockA{
\textit{Tejas Networks}\\
Bangalore, India \\
imteyaz@tejasnetworks.com}
\and
\IEEEauthorblockN{Jyotsna Bapat}
\IEEEauthorblockA{
\textit{IIIT Bangalore}\\
Bangalore, India \\
jbapat@iiitb.ac.in}
\and
\IEEEauthorblockN{Debabrata Das}
\IEEEauthorblockA{
\textit{IIIT Bangalore}\\
Bangalore, India\\
ddas@iiitb.ac.in}
}
\maketitle

\begin{abstract}
Failures in optical network backbone can lead to major disruption of internet data traffic. Hence, minimizing such failures is of paramount importance for the network operators. Even better, if the network failures can be predicted and preventive steps can be taken in advance to avoid any disruption in traffic. Various data driven and machine learning techniques have been proposed in literature for failure prediction. Most of these techniques need real time data from the networks and also need different monitors to measure key optical parameters. This means provision for failure prediction has to be available in network nodes, e.g., routers and network management systems. However, sometimes deployed networks do not have failure prediction built into their initial design but subsequently need arises for such mechanisms. For such systems, there are two key challenges. Firstly, statistics of failure distribution, data, etc., are not readily available. Secondly, major changes cannot be made to the network nodes which are already commercially deployed. This paper proposes a novel implementable non-intrusive failure prediction mechanism for deployed network nodes using information from log files of those devices. Numerical results show that the mechanism has near perfect accuracy in predicting failures of individual network nodes.
\end{abstract}

\begin{IEEEkeywords}
Failure prediction, optical networks, directed acyclic graph
\end{IEEEkeywords}

\section{Introduction}
Digital world (business, education, consumer, etc.) depends on access networks and backbone networks. The backbone network carries massive amount of data and interconnects all types of access networks to provide reliable connectivity among users, internet of things (IoT) devices, machines, clouds, etc. Most of backbone networks use optical communication, due to its strong advantage to support very high data rate. Backbone network consists of millions of routers/switches across the globe. Each router has multiple elements, and if one of them malfunctions, this may disconnect the router from rest of the network, leading to failure of critical global link incurring enormous disruption of data traffic, causing major loss to businesses and other activities. Hence, it is important to prevent failures using intelligent prediction mechanisms rather than a reactive approach of network recovery after a fault.

Several research proposals have been made on optical network failure prediction. \cite{cite_gaussian_classifier_heuristic_single_failure} applies Gaussian process classifier to detect single link failures. It also proposes heuristic to first identify the suspected links and then apply the classifier to identify the failed link. A support vector machine and double exponential smoothing based approach for network equipment failure prediction is described in \cite{cite_svn_time_series_net_node}.  \cite{cite_cellular_networks_bn} applies Bayesian networks to predict failures in cellular networks. Supervised learning based online and off-line techniques to predict link quality estimates in wireless sensor networks have been applied in \cite{cite_supervised_wsn}. \cite{cite_clustering_network_traffic_faults} compares three different data mining algorithms for network fault classification, namely, K-Means, Fuzzy C-Means, and Expectation Maximization, to suggest abnormal behavior in communication networks.

All the literatures explored above need large amount of data to be collected or simulated assuming certain distributions of various optical layer parameters. These techniques use various monitors for data collection. Sometimes, however, failure prediction is not built into the design of already deployed network systems, but subsequently need arises for such mechanisms. In such a scenario, two key challenges have to be dealt with. Firstly, statistics of failure distribution and relevant data are not readily available to directly apply conventional data driven approaches, such as, machine learning, Bayesian Networks, etc., \cite{cite_svn_time_series_net_node}\cite{cite_cellular_networks_bn}. Even if logs are available, extraction of useful statistics from all the historical data can be a time consuming process. Secondly, when such network nodes are commercially deployed in thousands neither major changes in software or hardware are possible, nor recommended to do so.

Logging mechanism in network nodes in form of text or binary files contain various information on variation of optical and system parameters, e.g., optical signal to noise ratio (OSNR), clock drift, etc. Log files also contain information on sequence of events over a period of time leading to different types of node failures.

This paper proposes a novel network node failure prediction using information from the log files without making any major changes to the deployed system. Using the log files to figure out the events leading to failures, a prediction mechanism is developed based on directed acyclic graph (DAG) and constructing two efficient standard data structures (section \ref{section_system_model}). The internal nodes in the DAG are the events and the leaf nodes are the failures. A directed edge exists from one event to the succeeding one, finally reaching a leaf failure node. A probability is defined for each node based on how far a node is from a possible failure (distance is defined in terms of number of hops to a failure). The nearer the node to a failure the higher the probability. The DAG and the associated data structures are constructed off-line from the old log files of similar errors as those the system is expected to predict. During the prediction phase, as events occur in real time the DAG is traversed through and failures are predicted in association with the data structures. Numerical results show near perfect failure prediction. To the best of knowledge of the authors, none of the previous work in the literature has proposed such a non-intrusive failure prediction mechanism for commercially deployed network nodes. 

There can be many possible implementations for such a network node failure prediction mechanism \cite{cite_gaussian_classifier_heuristic_single_failure}\cite{cite_svn_time_series_net_node}. Based on the amount of information available at this point of time, the following objectives are envisaged.
\begin{itemize}
\item Develop a non-intrusive system without making any changes to the commercially deployed system
\item Develop a quick first-cut solution, to reduce time-to-market
\item Use just enough information (available at this point of time) from the log files to develop the first-cut solution
\item Though, it is first-cut solution with a low number failures to predict, the system should evolve over time (for example, the DAG would extended as more log files are analyzed off-line to extract the appropriate event sequences leading to different failures)
\item Quick failure prediction phase
\item Use efficient standard data structures from the information in log files to have a quick prediction phase by ruling out invalid sequence of events, false-positives and converge on the valid transitions in the DAG
\item Efficient, safe and quick implementation of the prediction mechanism
\end{itemize}

The paper is organized as follows. The proposed idea and the system model is described in section \ref{section_system_model}. Results are discussed in section \ref{section_results}. Conclusions are drawn in section \ref{section_conclusion}.

\begin{figure}[ht]
\centering
\includegraphics[width=\columnwidth]{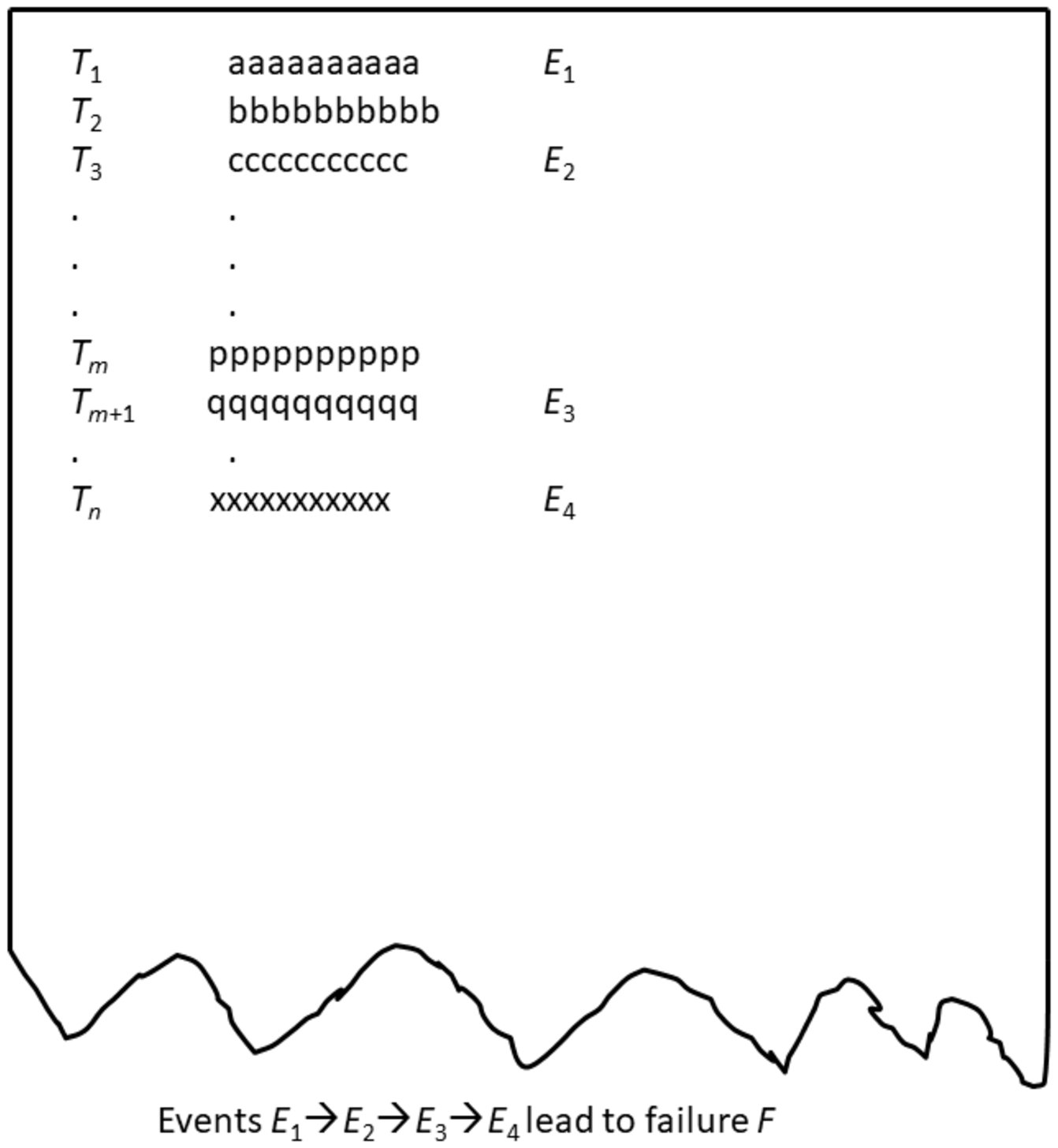}
\caption{Example log file}
\label{fig_sample_log_inkscape}
\end{figure}
\section{System Model}\label{section_system_model}
Most network systems have logging mechanism which helps in debugging malfunctionings and failures in the networks and their elements, e.g.,  routers, etc. Developers look at these logs to find out the sequence of events leading to a failure. Normally, logs contain information, such as, time stamp and associated texts, parameter values (e.g., clock drift), etc. (Fig. \ref{fig_sample_log_inkscape}).
The proposed idea uses these logs to build a fault prediction mechanism for network nodes, e.g., routers. This model looks for key information in the logs and converts them into sequences of events ($E_i$) and time ($T_i$), i.e., ($E_i, T_i$) tuples, leading to a failure ($i$ being the index). Consider section of an example log in Fig. \ref{fig_sample_log_inkscape}. Each network log contain important information in form of first two columns, namely, the time stamp $T_i$ and associated text and values of system parameters, e.g., clock drift, etc. When a failure occurs in a node, only the relevant information from the log files are designated as events  $E_i$ in the third column of the example log in Fig. \ref{fig_sample_log_inkscape}. For this example log file, to investigate a failure, information at times $T_1$, $T_3$, $T_{m+1}$ and $T_n$ are relevant and designated as events $E_1$, $E_2$, $E_3$ and $E_4$ respectively. Events can be, for example, clock drift beyond a certain threshold, rise in temperature above a certain limit, OSNR exceeding lower threshold, or a node not receiving signal from its peer, etc. It is assumed that the events are  spread out in time, otherwise it would be difficult to make a prediction and subsequent failure avoidance/prevention if they happen in quick succession. A DAG is created using these events for all the failures to be predicted by the system from old log files. Note that a couple of logs per failure are enough to identify the pattern of occurrences and designate certain information as events. This DAG and associated data structures (described below) are used for failure prediction.

\begin{figure}[ht]
\centering
\includegraphics[width=\columnwidth]{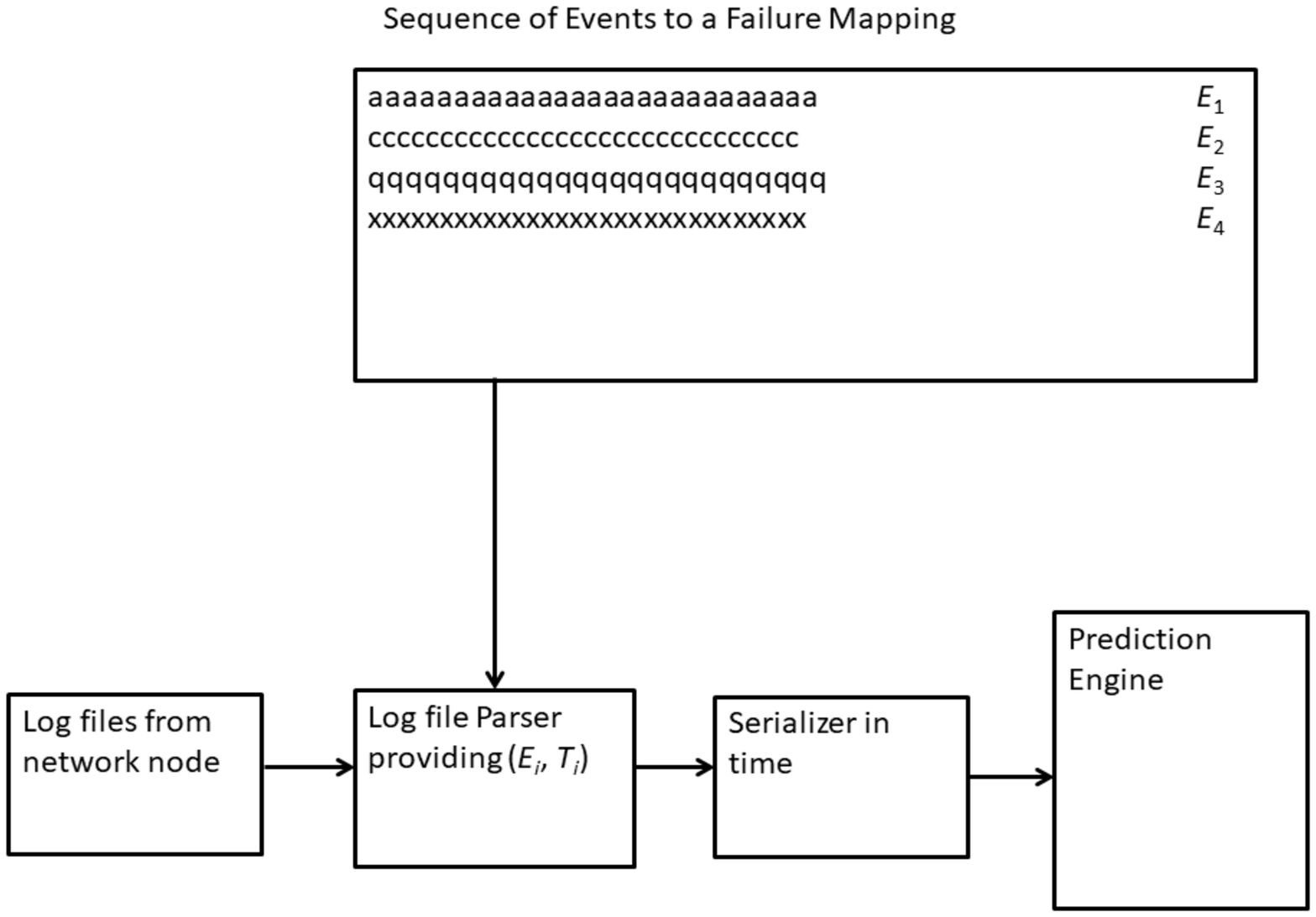}
\caption{Architecture of the proposed prediction mechanism}
\label{fig_architecture_inkscape}
\end{figure}

The overall architecture is shown in Fig. \ref{fig_architecture_inkscape}. During the prediction phase, log files are periodically read from the network device using remote copy, etc. The log file is then passed through as time based sliding window parser. This parser uses the mapping of text to event mapping (derived from the old log files) as explained above and outputs ($E_i, T_i$) tuples. A serializer sorts these tuples in time and passes the events to the DAG based prediction engine to check for failures.

Let there be maximum of $M$ events and $N$ failures in the system. This information is represented by a two dimensional array $E$  consisting $N$ rows and $M$ columns. Hence, each row vector represents the sequence of events leading to a failure. $E[i,j]$ is set to 1 if event $j$ happens leading to failure $i$, for $i=1,2,..,N$ and $j=1,2,..,M$, else the value is 0. It is assumed for any failure $i$ and $k,l \in j=1,2,..,M$, if $k \le l$ then $T_k \le T_l$, implying $E_k$ happens before $E_l$. $E$ can be filled up based on the experience of analyzing same failures from old log files.

The above array of  $E[i,j]$ consisting 1s and 0s is used as a general way to represent different kinds of failures. Events due to wrong configuration, one-off runtime events can be set easily. For periodic failures, two events may be required, one when the sequence starts and another when the same ends. For events based on variation of certain parameters, such as OSNR, different events may be set when they cross a lower or an upper threshold. Hence, this representation is general enough to handle most types of events leading to corresponding failures. Also, having 1s and 0s help in applying bitwise operators which helps in efficient implementation.

Using this matrix $E$, a DAG is created which is explained with an example below. Lets consider the matrix $E_{5,8}$ given below in (\ref{eqn_event_failure_table}).
\begin{equation}
E_{5,8} = \kbordermatrix{
    & E_1 & E_2 & E_3 & E_4 & E_5 & E_6 & E_7 & E_8\\
    F_1 & 1 & 0 & 0 & 0 & 1 & 1 & 0 & 0 \\
    F_2 & 0 & 0 & 1 & 1 & 0 & 1 & 1 & 0 \\
    F_3 & 0 & 1 & 0 & 0 & 1 & 0 & 1 & 1 \\
    F_4 & 0 & 0 & 0 & 0 & 1 & 1 & 0 & 0 \\
    F_5 & 1 & 0 & 0 & 0 & 1 & 0 & 0 & 1 \\
  }
  \label{eqn_event_failure_table}
\end{equation}
Events are represented along the columns of matrix $E_{5,8}$ and failures along rows. Failure $F_1$ happens in the sequence $E_1 \rightarrow E_5 \rightarrow E_6$. Other failures happen in same way.

\begin{figure}[ht]
\centering
\includegraphics[width=\columnwidth]{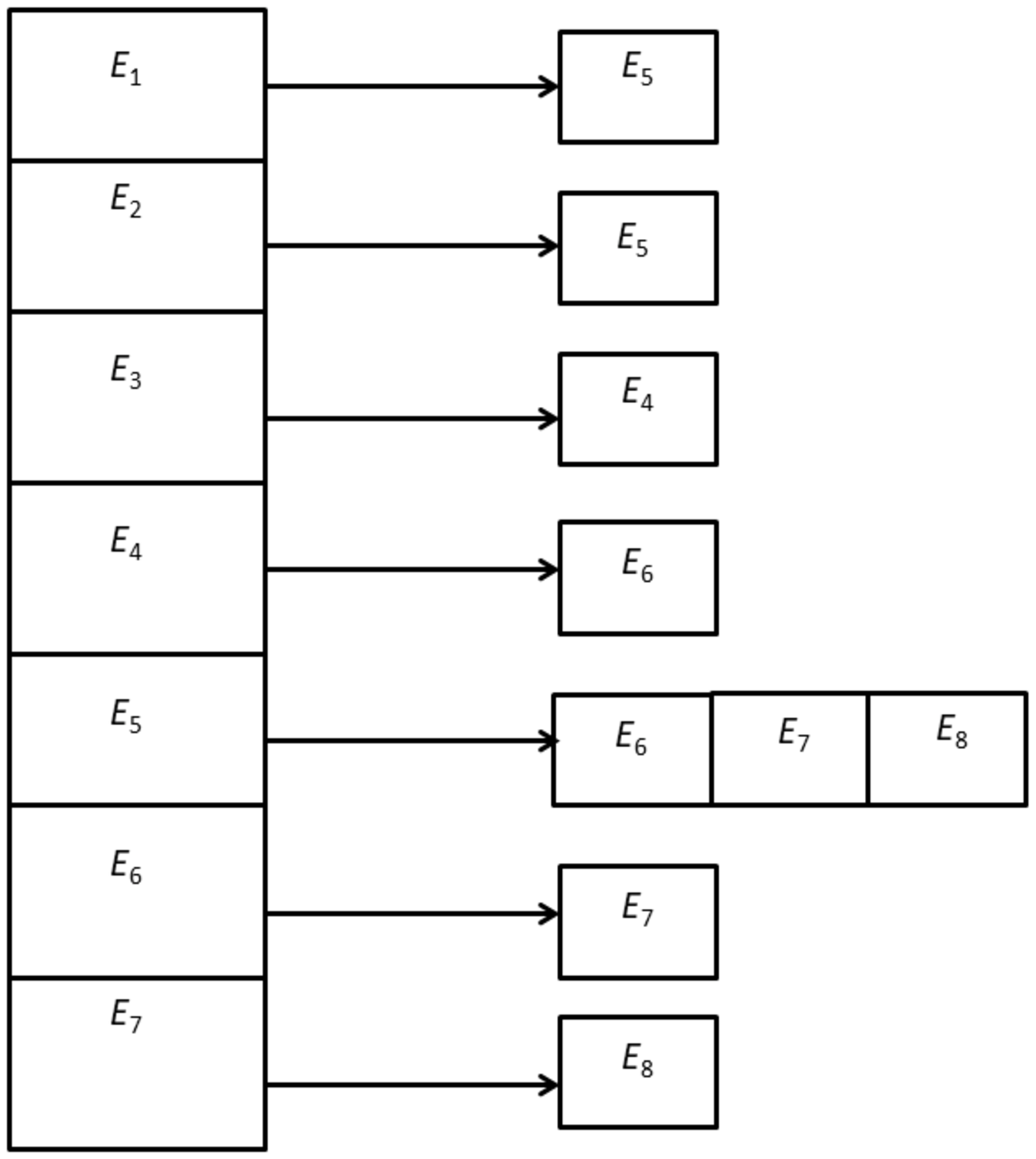}
\caption{DAG data structure}
\label{fig_DAG_data_structure_inkscape}
\end{figure}

\begin{figure}[ht]
\centering
\includegraphics[width=\columnwidth]{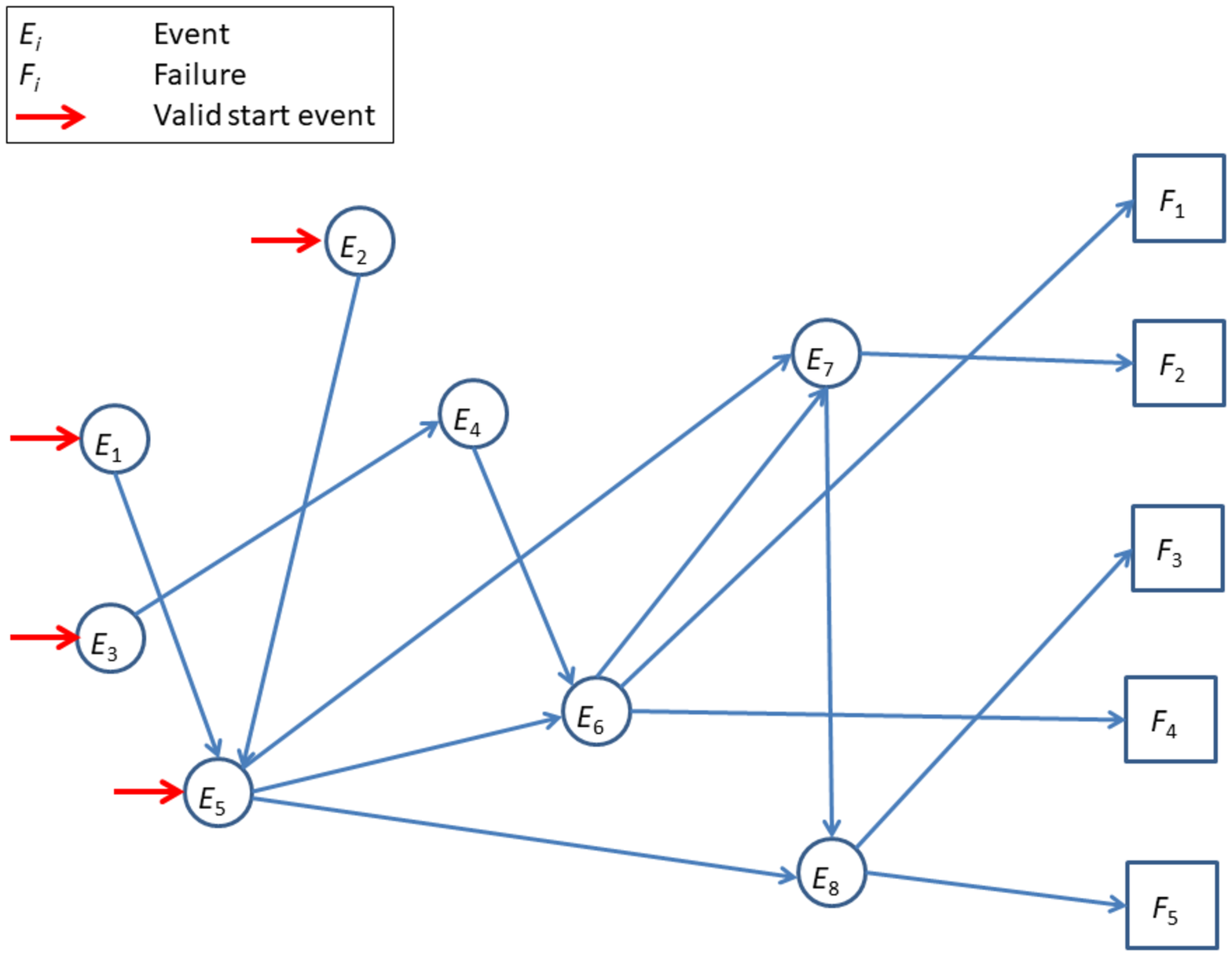}
\caption{DAG with 8 events and 5 failures}
\label{fig_DAG-8-events-5-failures_inkscape}
\end{figure}

Generally, each of the events $E_j$ and failures $F_i$, for $i=1,2,..,N$ and $j=1,2,..,M$, are the nodes of the DAG and the sequence of events are linked with edges. For the example matrix in (\ref{eqn_event_failure_table}), the DAG is shown in Fig \ref{fig_DAG-8-events-5-failures_inkscape}. Note that for $M$ events there are $2^M - 1$ (leaving out the one with all zeros) sequences are possible. This DAG can be constructed once while initializing the system and the same can be used during the online failure prediction phase. The data structure for the DAG is an array of nodes (vertically) and each of edges to it neighbours is a list (horizontally) as shown in Fig. \ref{fig_DAG_data_structure_inkscape}. The DAG is constructed using information from off-line analysis by the developers. Real time logs cannot be used to build the DAG.

Since, the only information available is the sequence events leading to a failure, conventional data driven approaches, such as, Bayesian Networks \cite{cite_cellular_networks_bn}, etc., cannot be applied. To circumvent this problem and still have working solution, probability of failure is defined by how much a node is closer to a failure. For example, probability of $F_1$ is higher when event $E_5$ happens compared to when $E_1$ occurs. However, probability of $F_1$ and $F_5$ are equal when $E_1$ occurs, since the distances from both failures are same (3 hops) by (\ref{eqn_event_failure_table}).

Since, the DAG and the matrix $E$ are static information, these can be used to construct a hop matrix $H_{8,5}$ in (\ref{eqn_hop_matrix_table}) which will contain the number of hops to a failure. For example, $E_1$ has 3 hops to failures $F_1$ and $F_5$. Note that $F_1$ and $F_5$ are the only possible failures according to (\ref{eqn_event_failure_table}) when $E_1$ occurs, $F_2$, $F_3$ and $F_4$ are ruled out. Similarly, $E_4$ has 3 hops for $F_2$, event $E_5$ has 2 hops to failures $F_1$, $F_4$ and $F_5$, and 3 to $F_3$.


\begin{equation}
H_{8,5} = \kbordermatrix{
    & F_1 & F_2 & F_3 & F_4 & F_5\\
    E_1 & 3 & 0 & 0 & 0 & 3 \\
    E_2 & 0 & 0 & 4 & 0 & 0 \\
    E_3 & 0 & 4 & 0 & 0 & 0 \\
    E_4 & 0 & 3 & 0 & 0 & 0 \\
    E_5 & 2 & 0 & 3 & 2 & 2 \\
    E_6 & 1 & 2 & 0 & 1 & 0 \\
    E_7 & 0 & 1 & 2 & 0 & 0 \\
    E_8 & 0 & 0 & 1 & 0 & 1
  }
  \label{eqn_hop_matrix_table}
\end{equation}

Probability of a failure $F_i$ from any node (event $E_j$) in DAG is defined in (\ref{eqn_failure_prob}). The essential intuition behind this definition is to increase the probability exponentially so that the failure avoidance/prevention mechanism can be triggered at required threshold value. Note that this definition holds good only for the hop matrix (\ref{eqn_hop_matrix_table}) and not generalized to any number of hops. In future, the function will be generalized to any number of nodes using concepts like network diameter, etc.
\begin{equation}\label{eqn_failure_prob}
P^{(E_j)}_{F_i} = \frac{100 - e^{H_{8,5}(j,i)}}{100}
\end{equation}

\subsection{Failure prediction}
Prediction process begins with the valid start events. In Fig. \ref{fig_DAG-8-events-5-failures_inkscape}, $E_1$, $E_2$, $E_3$ and $E_5$ are the valid start events (marked in red), rest are not.
Once the DAG is constructed off-line, failure prediction starts with its traversal, as the events occur in real time, and some invalid sequences are filtered out (from the total of $2^M - 1$ possible sequences) for which there are no edges between nodes. For example, $E_8$ cannot succeed $E_6$.

At each hop, all the failures reachable from that node are predicted using (\ref{eqn_hop_matrix_table}). This is not very efficient.
One important enhancement that can be made is to prune the number of failures that it is trying to predict. For example, considering $F_1$ and $F_5$, if $E_1$ occurs then the proposed mechanism will predict both the failures with equal probability using (\ref{eqn_event_failure_table}), (\ref{eqn_hop_matrix_table}) and (\ref{eqn_failure_prob}). It can be observed that the sequence of events for $F_1$ ($E_1 \rightarrow E_5 \rightarrow E_6$) and $F_5$ ($E_1 \rightarrow  E_5 \rightarrow E_8$) are different. For this purpose, it is essential to keep track of the path to reach a node in the DAG to predict failures accurately.

To uniquely identify the path, a heuristic is developed. A global event bit-mask is constructed when an event is received (starting with the valid first events in Fig. \ref{fig_DAG-8-events-5-failures_inkscape}). This event bit-mask is initialized to zero. Bitwise AND operations is performed with the event bit-mask and each row of matrix $E_{5,8}$ (\ref{eqn_event_failure_table}). Those output of the AND operations that do not have the bits set to 1 are left out of contention and ones remaining are probable failures. These steps are continued until the correct failure is predicted. Note that this enhanced procedure, to uniquely identify the path to a failure, needs to be performed only when there is a valid transition in the DAG. Also, if there is single transition from the present state in DAG, this procedure is not necessary. This working of the event bit-mask is explained in detail in the results section (section \ref{section_results}). This process to identification of failures has some additional computational overhead of the order of $O(N)$, assuming that the bitwise AND operation of the two vectors takes one time step.

Apparently, it may appear that a depth first search (DFS) may help in arriving at a failure. But, DFS is more useful and applied in exploratory search of the entire graph. Here, this is not the case because the proposed idea looks for the correct path to a failure with static information already available in the data structures. Also, DFS has higher time complexity compared to the approach taken here.

An important implementation point to be noted here is that an event can either be a valid starting point for one failure and an intermediate one for a different failure. For example, $E_5$ is the starting point for $F_4$ and an intermediate one for $F_1$. Hence, when $E_5$ occurs it is important to find out whether it is a beginning of a new sequence of events for $F_4$ or an intermediate one for $F_1$. Hence, concurrent invocation of the prediction engine for multiple failures is necessary. This can be done by making the implementation re-entrant which can be invoked by multiple threads concurrently. Each invocation of the prediction engine with try predict an independent failure.

\section{Results and Discussion}\label{section_results}
This section discusses the results obtained by applying the network node failure prediction proposed above. For this purpose, the same DAG in Fig. \ref{fig_DAG-8-events-5-failures_inkscape} is used to demonstrate the proposal. The model is implemented using \verb|python numpy| libraries.
\subsection{DAG based invalid sequence of detection}
If there are 8 events, then there can be $2^8 - 1$ sequences possible, out of which only 5 are considered valid by \ref{eqn_event_failure_table}. DAG helps in filtering out some of the invalid sequences. Lets consider an invalid sequence $(E_1 = 1, E_2 = 0, E_3 = 0, E_4 = 0, E_5 = 1, E_6 = 1, E_7 = 0, E_8 = 1)$. If this sequence of events are propagated through the DAG, it is evident there is no transition from $E_6$ to $E_8$. Hence, an invalid sequence is declared. Thus, sequence of events which do not have transitions (edges) in the DAG get discarded in this step. However, the invalid sequence $(E_1 = 1, E_2 = 0, E_3 = 0, E_4 = 0, E_5 = 1, E_6 = 1, E_7 = 1, E_8 = 1)$ will predict failures $F_3$ and $F_5$. To circumvent these false-positive cases, the following steps are necessary.
\subsection{Sequence of events leading to failure}
\begin{figure}[ht]
\centering
\includegraphics[width=\columnwidth]{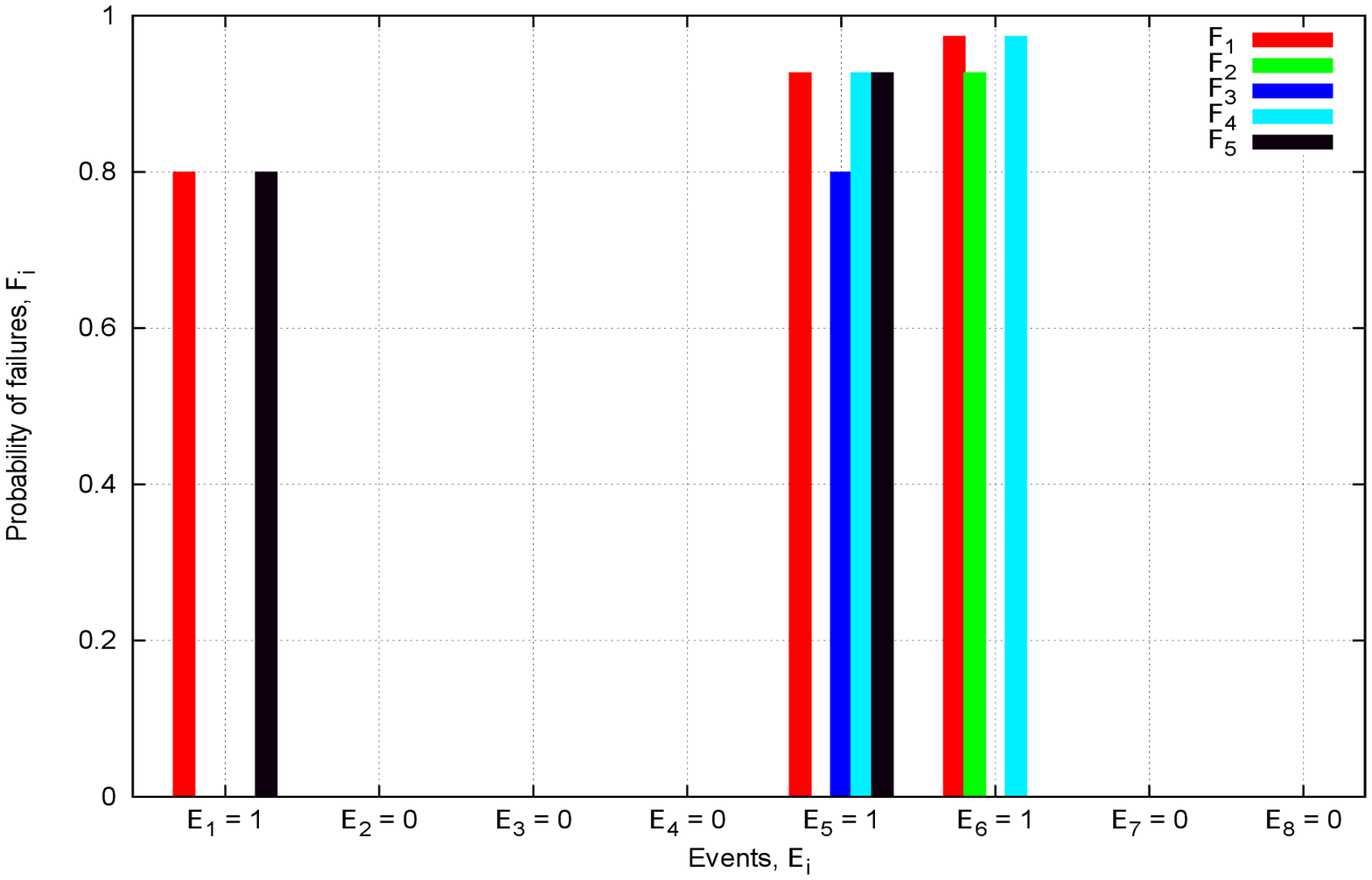}
\caption{Sequence of events $E_i$ leading to failure $F_1$}
\label{fig_plot_events_to_failure_1_inkscape}
\end{figure}

Fig. \ref{fig_plot_events_to_failure_1_inkscape} steps through the failure detection for $F_1$ when the relevant events occur. Along \emph{x}-axis the sequence of events $E_i$, $i$ = 1, 2,.., 8, in time leading to $F_1$ are shown. Note that $E_i$ = 1 means the event has occurred and 0 otherwise, and $F_1$ occurs when the following ordered tuple of events happen $(E_1 = 1, E_2 = 0, E_3 = 0, E_4 = 0, E_5 = 1, E_6 = 1, E_7 = 0, E_8 = 0)$. The probabilities (\ref{eqn_failure_prob}) of relevant failures are shown along \emph{y}-axis. When $E_1$ occurs the model predicts $F_1$ and $F_5$ using the hop matrix (\ref{eqn_hop_matrix_table}) which rules out other failures. Subsequently, the model predicts $F_1$, $F_3$, $F_4$ and $F_5$ when $E_5$ occurs, and $F_1$, $F_2$, $F_4$ when $E_6$ occurs.

The above results have some redundancy since the model predicts two additional failures $F_2$ and $F_4$ along with the correct one $F_1$ when $E_6$ occurs. Next section applies the enhancement with event bit-mask to optimize the prediction.
\subsection{Sequence of events leading to failure with enhancements}
\begin{figure}[ht]
\centering
\includegraphics[width=\columnwidth]{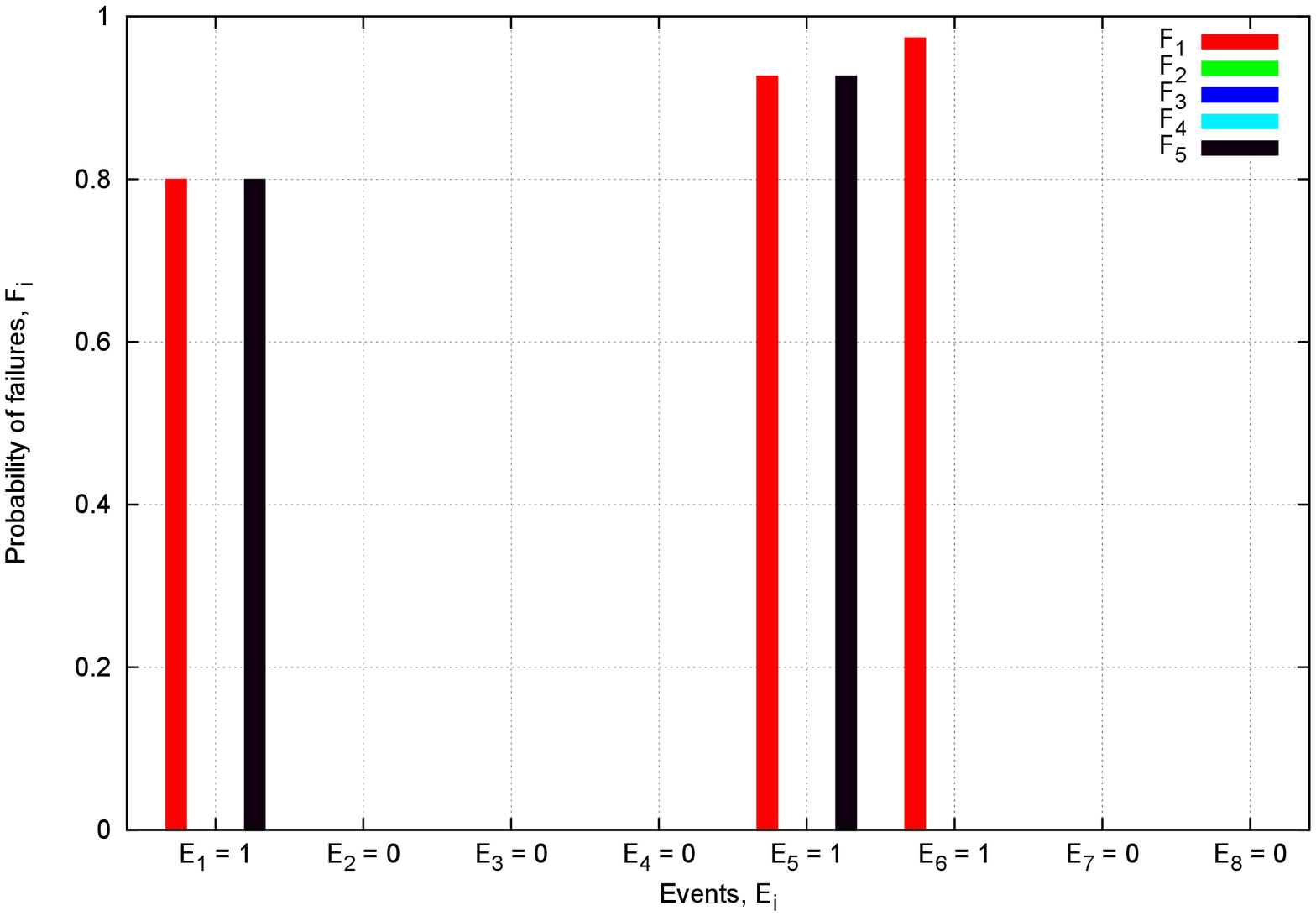}
\caption{Sequence of events $E_i$ leading to failure $F_1$ with enhancements}
\label{fig_plot_events_to_failure_1_w_enh_inkscape}
\end{figure}

Fig. \ref{fig_plot_events_to_failure_1_w_enh_inkscape} predicts $F_1$ and $F_5$ when $E_1$ occurs and constructs a event bit-mask [1, 0, 0, 0, 0, 0, 0, 0]. Since, $E_1$ triggers valid transition in the DAG, a bitwise AND operation is performed with rows of (\ref{eqn_event_failure_table}) and rules out failures $F_2$, $F_3$ and $F_4$. When event $E_5$ (a valid transition in DAG) occurs the mask is updated to [1, 0, 0, 0, 1, 0, 0, 0]. Again, when this mask is applied to $F_1$ and $F_5$ in (\ref{eqn_event_failure_table}), both are retained ($F_2$, $F_3$, $F_4$ are already ruled out in the previous step). Finally, event $E_6$ leads to the updated event bit-mask as [1, 0, 0, 0, 1, 1, 0, 0]. This mask is applied to $F_1$ and $F_5$, the latter is ruled out. Hence, $F_1$ is the only failure predicted which is as expected.
\section{Conclusion}\label{section_conclusion}
Failure prediction in optical backbone network is extremely important to avoid large scale disruption of data traffic. However, such prediction mechanism is sometimes not built into the network nodes at design time and subsequently the need arises to have one. This paper presented an implementable non-intrusive failure prediction mechanism in network nodes making use of existing log files. The proposed idea constructs a DAG and other associated data structures from key events in the log files resulting in a failure. Numerical results show that the proposed idea is able to predict the failures in a near perfect way.

Future work will hinge on extending the model to higher number of nodes, and performance analysis after integration with the commercially deployed network.

\section*{Acknowledgment}

The authors would like to thank Tejas Networks for sponsoring this research project.

\bibliographystyle{IEEEtran}
\bibliography{IEEEabrv,net_failure_pred}

\end{document}